\documentclass[aps,prl,twocolumn,showpacs,preprintnumbers]{revtex4-1}

\usepackage{psfrag,graphicx}
\usepackage{dcolumn}
\usepackage{amsmath,amssymb}
\usepackage{bm}
\usepackage{amsfonts,amssymb,amsmath}        
\usepackage{epstopdf}

\newcommand{\be}{\begin{equation}}
\newcommand{\ee}{\end{equation}}
\newcommand{\bq}{\begin{eqnarray}}
\newcommand{\eq}{\end{eqnarray}}

\begin{document}

\title{Novel Topological Phases and Self-Correcting Memories in Interacting Anyon Systems}
\author{James R. Wootton}
\affiliation{Department of Physics, University of Basel, Klingelbergstrasse 82, CH-4056 Basel, Switzerland}

\begin{abstract}

Recent studies have shown that topological models with interacting anyonic quasiparticles can be used as self-correcting quantum memories. Here we study the behaviour of these models at thermal equilibrium. It is found that the interactions allow topological order to exist at finite temperature, not only in an extension of the ground state phase but also in a novel form of topologically ordered phase. Both phases are found to support self-correction in all models considered, and the transition between them corresponds to a change in the scaling of memory lifetime with system size.

\end{abstract}


\maketitle

\textit{Introduction:-} Topologically ordered systems have recently generated a lot of interest in both the fields of condensed matter and quantum information. Particular attention has been given to the possibility of using topological systems to create self-correcting quantum memories, with much progress made toward determining the kinds of systems which can and cannot support them \cite{bacon,alicki1,ortiz,kaycol,alicki2,bravter,alicki3,kay2,pasta,haah1,haah2,michnicki}. See \cite{woottonrev} for a recent review. It has been shown that systems of interacting anyonic quasiparticles in particular are an important and effective candidate for a realistic solution \cite{hamma,chesi,fabio,beat,hutter,fahut,becker}. It is therefore necessary to fully understand such systems, including their properties at thermal equilibrium, the phases that are present and how these relate to the behaviour of the quantum memory. It is such an equilibrium study of interacting anyon systems that we perform here, using the anyonic topological entropy \cite{wootton}. We find that anyonic interactions support the existence of two distinct topological ordered phases at finite temperature, both of which correspond to self-correcting quantum memories.

It should be noted that since anyonic interactions do not occur naturally, they must be engineered as effective interactions in a more complex physical system. In this study we wish to study only the nature of anyonic interactions, and so consider the equilibrium distribution of the effective Hamiltonian alone.

\begin{figure}[t]
\begin{center}
{\includegraphics[width=8.5cm]{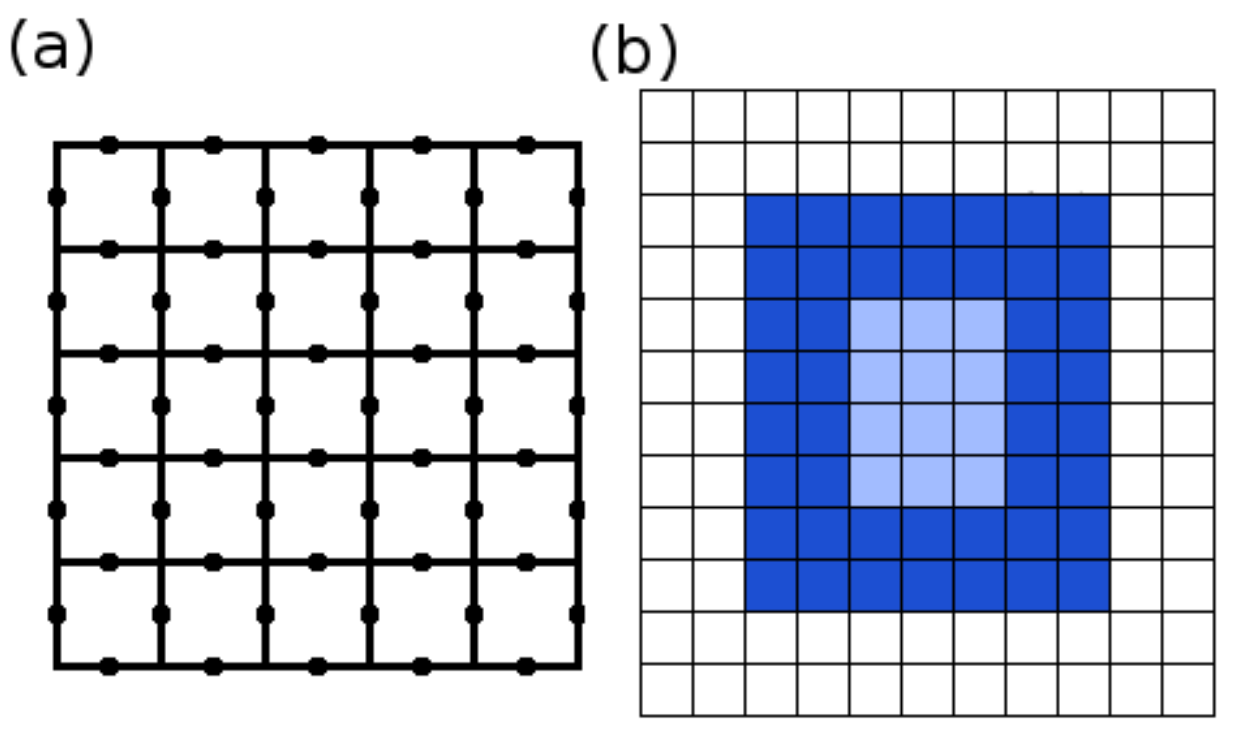}}
\caption{\label{fig1} (a) The spin lattice on which an periodic $L \times L$ toric code is defined. A spin-$1/2$ particle resides on each edge. (b) The partition of the plaquettes used for the calculation of the anyonic topological entropy. The region $R_a$ is shown in light blue, $R_b$ in dark blue and $R_c$ in white. The thickness of $R_b$ is assumed to be of the same order as that of $R_a$. A corresponding partition is used for the vertices}
\end{center}
\end{figure}

\textit{Model of interacting anyons:-} Proposals for interacting anyon quantum memories have focussed on the surface codes, and the toric code in particular \cite{wegner,dennis,kitaev}. This is defined on an $L \times L$ square lattice with periodic boundary conditions and a spin-$1/2$ on each edge, as shown in Fig. \ref{fig1}(a). Operators are defined for each plaquette, $p$, and vertex, $v$, of the lattice as follows,
\begin{equation} \nonumber
n_p = \frac{1-\prod_{i \in p} \sigma^z_i}{2}, \,\,\, n_v = \frac{1-\prod_{i \in v} \sigma^x_i}{2}.
\end{equation}
Here $i \in p$ denotes that the sum is over all spins around the plaquette $p$, etc. Each $n_p$ operator is a projector onto the subspace of states for which an $m$ anyon is present on the plaquette $p$. Similarly the $n_v$ are projectors for $e$ anyons on vertices.

In this work, we will primarily consider three forms of interaction for the anyons. These act on the plaquette and vertex anyons independently, but equivalently. Their form, in terms of the plaquettes, is,
\bq \label{non}
H_0 &=& J \sum_p n_p, \\ \label{c}
H_C &=& J \sum_p n_p + A \sum_{p} \sum_{p' \neq p} n_p n_{p'} {r_{pp'}^{-\alpha}}, \\ \label{h}
H_H &=& J \sum_p n_p + A \sum_{p} \sum_{p' \neq p} n_p n_{p'} \, \ln r_{pp'}.
\eq
Here $J$ and $A$ are arbitrary coupling strengths, which are positive constants unless otherwise specified, and $r_{pp'}$ is the Euclidean distance between the plaquettes $p$ and $p'$. The form for vertices is found through the substitution $p \rightarrow v$. $H_0$ is the toric code Hamiltonian without anyonic interactions \cite{kitaev}, whereas $H_C$ and $H_H$ are those with the interactions proposed and studied by Chesi et al. \cite{chesi,fabio,beat,hutter} and Hamma et al. \cite{hamma}, respectively.

\textit{Range of topological correlations:-} The correlations of topologically ordered systems are unique in that they can be detected only by correlation functions defined around closed loops  \cite{levwen,hastings}. For the unperturbed toric code at zero temperature, these correlation functions are simply the product of all plaquette or vertex operators inside the loop. However, in general they can become quite complex `fattened' loop operators  \cite{levwen,hastings}. Nevertheless, it is possible to detect the existence of these topological correlations using entropic effects, even if the form of the correlation function is not known. It was this principle that was used to derive the topological entropy of Levin and Wen \cite{levwen} as well as the anyonic topological entropy \cite{wootton}. These are equivalent to each other, and to the formulation of Kitaev and Preskill \cite{kitpres}.

In this work the state of the system will be most naturally expressed in terms of the anyon occupancies, rather than the underlying spins. As such, when attempting to determine whether the loop correlations exist or not at various length scales, it is most convenient to use the anyonic topological entropy, $\Gamma$. This is calculated by first partitioning the plaquettes and vertices (not the spins, as in the alternative formulations) of the lattice into the three regions $R_a$, $R_b$ and $R_c$, shown in Fig. \ref{fig1}(b). The quantity is then defined as $\Gamma = 2 \log 2 - \Gamma'$, where $\Gamma'$ is the amount of information concerning the net occupation of $R_a$ that cannot be determined from the anyon configuration of $R_b$. When there are no spatial correlations between the positions of anyons, as will be considered in the bulk of this work, $R_b$ is able to deduce nothing about $R_a$. This means $\Gamma'$ simply equal to the entropy of the probability distribution for the net occupancy of $R_a$, and so
\be \label{Gamma}
\Gamma = 2 \log 2 - S(\pi_P) - S(\pi_V),
\ee
where $\pi_P$ ($\pi_V$) is the probability that there is no net $m$ ($e$) in the plaquettes (vertices) of the region $R_a$, and $S(\pi) = -\pi \log \pi - (1-\pi) \log (1-\pi)$ is the Shannon entropy. This entropy is zero when there is no randomness in the net occupancies and $\log 2$ when they are fully random. We therefore obtain $\Gamma= 2 \log 2$ when the loop correlations around $R_b$ are fully present and $\Gamma=0$ when they are not present at all. An intermediary value implies that the correlations exist, but are of lower fidelity that the ideal case since the so-called `fattened' loop operators cannot fully correct for the net occupancies inside the loop \cite{hastings,wootton}. Note that the mixed state generalization of the Kitaev and Preskill topological entropy is given by the same expression in \cite{iblisdir}. The results obtained in this work could therefore be equivalent obtained using this measure.

States are usually considered to be topologically ordered if the loop correlations exist for arbitrarily large fidelity for arbitrarily large loops. This means $\Gamma \rightarrow 2 \log 2$ as $L \rightarrow \infty$ when $l/L = O(1)$. However, to gain more insight into the behaviour of the topological correlations, in this work we will consider how their strength varies with the size of the loop, $l$.

Let us first consider the non-interacting case, $H_0$. Though the system is topologically ordered and forms a stable quantum memory at zero temperature \cite{bravyi,vidal,James,Cyril,kay1}, it becomes disordered \cite{claudios,iblisdir} and unstable \cite{alicki2} at finite temperature. Clearly the Hamiltonian acts independently on each plaquette (apart from a global parity constraint, whose effects are only evident when the joint state of almost all plaquettes is considered). The probability of an anyon on each plaquette can then be found from the Boltzmann distribution,
\be \label{boltzmann}
\rho = (e^{J \beta} + 1)^{-1},
\ee
where $\beta$ is the inverse temperature. The same probability will apply to vertex anyons. The fusion rules of the anyons mean that there is no net $m$ ($e$) anyon within $R_a$ if there are an even number of $m$ ($e$) anyons. The probability of this is
\be \nonumber
\pi_P(l) = \pi_V(l) = \frac{1 + (1-2\rho)^{l^2} }{2}.
\ee
And so,
\be \nonumber
\Gamma(l) = 2 [\log 2 - S(\pi_P(l))],
\ee
Clearly $\pi_P(l) \approx 1$ and $\Gamma(l) \approx 2 \log 2$ for small $l$, because the probability of any anyon existing within the area is small. However, as $l$ increases, $\pi_P(l)$ will tend to $1/2$ and $\Gamma(l)$ to $0$. The topological correlations are therefore only present at a short range. To quantify this range, let us say that the correlations persist while $\Gamma(l)$ is
within a tolerance $\Delta$ of its maximum value. This implies that $\pi_P(l)$ must be within a corresponding tolerance $\delta$ of zero. The range $\lambda$ of the correlations can then be defined as that at which $\pi_P(\lambda) = \delta$. Rearrangement of the above leads to,
\be \label{eq}
\lambda^2 = \frac{\ln(1-2\delta)}{\ln(1-2\rho)}.
\ee
Since $\rho$ does not vary with system size, we find that $\lambda = O(L^0)$. The range of the topological correlations is finite and constant with respect to system size for any finite temperature. This is exactly the behaviour to be expected from states that are not topologically ordered, and so we confirm the known result that the state at finite temperature is not topologically ordered when there are no anyonic interactions \cite{claudios,iblisdir}.

\textit{Power law potential:-} We will now consider the power law potential given in the Hamiltonian $H_C$ of Eq. (\ref{c}). For any finite $\alpha$, the energy of an anyon configuration depends on the distances between the anyons. Correlations will therefore exist between their relative positions. The extreme case of this, obtained for the lowest energy configuration for any number of anyons, is that the anyons form an ordered lattice. If such an anyonic crystal is stable, the state will be topologically ordered. This is true even if the position and orientation of the lattice is random, since measurement of the crystal in the region $R_b$ is sufficient to deduce its structure in (and hence the net occupancy of) $R_a$. However, the level of stability required to satisfy this condition is very high. For example, it requires that no anyons are displaced a distance $O(L)$ from their proper position. Otherwise these could leave or enter the region $R_a$ without leaving a trace in $R_b$, preventing the net occupancy of the former from being deduced by the latter. For any $\alpha \geq 0$, the energy cost required for an anyon to escape from its proper position to infinity is finite. Hence, at any finite temperature, a fraction of the anyons will indeed be dislocated in such a manner, destroying the topological order. This is in addition to many other smaller dislocations and defects that will further prevent the lattice structure being used to determine the net occupancy of $R_a$ from the configuration within $R_b$. The correlations between anyons are therefore far too weak to have a significant effect on the topological order of the system, and can be ignored for all non-negative $\alpha$. This allows us to adopt a mean field treatment, and to continue using the relations derived above for the uncorrelated case.

We first consider interactions of the form $A > 0$ and $\alpha > 2$. Assuming that anyons are evenly spread out with a density $\rho$, and taking the continuum limit, we see the additional energy per anyon due to the interaction potential is,
\be \nonumber
\epsilon = \int_{r=1}^\infty \frac{A}{r^\alpha} \, 2 \pi r \, \rho \, dr = \frac{2 \pi A \rho}{\alpha-2}.
\ee
Clearly $\epsilon = O(L^0)$. The interaction therefore results only in an effective $O(1)$ addition to the coupling $J$. This again leads to $\lambda = O(L^0)$ and so demonstrates that thermal states for such interactions are not topologically ordered.

For the case of $0 \leq \alpha < 2$, the mean field density of anyons was found in \cite{chesi} to be well approximated by,
\be \label{rho}
\rho = \frac{2-\alpha}{c_{\alpha}\beta A} \frac{\ln L}{L^{2-\alpha}}.
\ee
Here $c_{\alpha}$ is an $O(1)$ parameter that depends on $\alpha$ alone. Note that the interaction is constant when $\alpha=0$,
leading to no spacial correlations and making the mean field treatment exact. From this expression, Eq. (\ref{eq}) may be used to determine $\lambda$. For large $L$, $\rho$ will be small enough to make the approximation $\ln(1-2\rho) \approx - 2 \rho$. Hence,
\be \label{lambda}
\lambda^2 = \frac{c_{\alpha}\beta A \ln(1-2\delta) }{\alpha-2}\frac{L^{2-\alpha}}{\ln L}, \,\,\, \therefore \,\,\, \lambda = O(L^{1-\alpha/2}).
\ee
Here we see that $\lambda$ increases with $L$ according to a power law. However, since the scaling is sub-linear, it does not result in $\Gamma \rightarrow 2 \log 2$ as $L \rightarrow \infty$ when $l / L = O(1)$. This is a novel form of behaviour. Like the topologically ordered ground state of the model, the range of topological correlations diverges with system size. However, unlike the ground state, the correlations never permeate the entire system. The former fact demonstrates that such states are topologically ordered, but the latter shows that the topologically ordered phase is distinct from that of the ground state. Henceforth we will refer to such states as `weakly topologically ordered'. States for which $\Gamma \rightarrow 2 \log 2$ as $L \rightarrow \infty$ when $l/L = O(1)$ will be referred to as `strongly topologically ordered'.

The case of $\alpha < 0$ is unphysical, since it is hard to imagine a physical system that could induce such interactions. However, since it can be straightforwardly analysed, we will consider it here for completeness. In the case of $\alpha < 0$, the potential becomes attractive and scales as $A r^{|\alpha |}$. This results in an exponentially decaying bound of $O(e^{-\beta A r^{|\alpha |}})$ for the distance between any pair of anyons, and so all anyons will be confined to a single area of finite radius $d$. The net occupancy of $R_a$ can only not be determined from the configuration of $R_b$ if the single cluster of anyons spans the width of the annulus. Since this will not occur so long as $l > d$, the thermal states for these interactions are strongly topologically ordered.

It is interesting to compare the behaviour of $\lambda$ to the lifetime of the quantum memory, as determined in previous work \cite{chesi,abbas}. The $\alpha>2$ regime with no topological order has been shown to have a lifetime of $\tau = O(L^0)$, and so does not benefit from topological protection against thermal errors. The weakly topological ordered $0 \leq \alpha < 2$ regime has $\tau = O(L^{2-\alpha})$, and hence has a lifetime that scales polynomially with the system size. For $\alpha < 0$, the exponentially decaying probability of any anyon existing outside of a finite area means that it will take an amount of time exponential with $L^{|\alpha |}$ for any anyon to escape and propagate along the $O(L)$ distance required cause a logical error. The onset of the strongly topologically ordered phase therefore increases lifetime from the polynomial scaling of the weakly topologically ordered case, to exponential scaling.

\textit{Logarithmic potential:-} Like the $\alpha<0$ potential above, the logarithmic potential of Eq. (\ref{h}) is attractive and divergent for $A>0$. It will encourage anyons to collapse around a point, and can induce strong spatial correlations between anyons that cannot be ignored.

To analyse the effects of the interactions in this model, it is instructive to first consider the case for which the anyon number is fixed. For the case of two anyons, one of which can be chosen to reside at the origin, the (unnormalized) probability that the other is a distance $r$ away is $e^{-\beta A \ln r} = r^{-\beta A}$. Taking the continuum limit, the probability that this anyon resides within a radius $R$ is,
\be \label{int}
p(R) = \frac{ \int_1^R r^{-\beta A} \, 2\pi r \, dr     }{ \int_1^L r^{-\beta A} \, 2\pi r \, dr } = \frac{R^{2-\beta A}-1}{L^{2-\beta A}-1}.
\ee
Note that circular boundary conditions were taken in the denominator rather than square for simplicity. Nevertheless, the result will have the same scaling with $L$ and so be equivalent for our purposes. The behaviour of $p(R)$ clearly has a critical temperature at $T_c = A/2$. For $T<T_c$, the exponents in the r.h.s. of Eq. (\ref{int}) will be negative. This yields $p(R) =1-O(R^{-|2-\beta A|})$ as $L\rightarrow\infty$, and so there is an arbitrarily high probability of the anyons being separated only by a finite distance. For $T>T_c$ the exponents are positive, leading to $p(R)=O \left((R/L)^{|2-\beta A|} \right)$. This means $p(R)=O(1)$ when $R/L = O(1)$, even as $L \rightarrow \infty$. The spatial correlations are therefore weak, with each anyon allowed to explore the entire system independently of the other. As noted in \cite{hamma,hamref1,hamref2}, any collection of $N$ anyons will similarly have a critical temperature $T_c (N) = N A/4$, below which they are confined around a point and above which they are deconfined.

For the full problem, where the total anyon number is not fixed, the total thermal state is a statistical mixture of the thermal states for different fixed numbers of anyons. Since $N=2$ is the smallest number of anyons possible in the toric code, the temperature $T_c = A/2$ is that under which any number of anyons is confined. The thermal state is therefore strongly topologically ordered below this. For any temperature $T_c(N) < T < T_c(N+1)$ above this point, all states of $N$ or less anyons will be deconfined (and so have a divergent partition function) and all those with more will be confined (a convergent partition function). The former will therefore dominate the latter in the total thermal state, and so the average number of anyons will be $\leq N$. Since this means only a finite number of anyons will be present on average at any finite temperature, the relation of Eq. (\ref{eq}) yields $\lambda = O(L)$. The state is therefore weakly topologically ordered, with the highest possible scaling of $\lambda$ with $L$. Any higher scaling would result in the topological correlations permeating the entire system, making it strongly topologically ordered.

It is argued in \cite{hamma} that the lifetime of the memory with such interactions scales as $\tau = O(L^{\beta A})$ when $T<T_c$. No argument for the case of $T>T_c$ was given, however the behaviour here can be easily seen by noting that the interactions cause only a finite number of anyons to be present, but have only a weak attractive effect at such temperatures. The lifetime should then scale similarly to the case of a finite number of anyons in a model with no interactions, i.e. $\tau = O(L^2)$ \cite{chesi,abbas}. Since $T = T_c$ implies $\beta A=2$, this means that the exponent of $L$ decreases with temperature  while in the topologically ordered phase, until it reaches the value of $2$. The state then transitions to the weakly topological ordered phase and the exponent remains at this minimum value.

Note that since a single anyon can exist on the planar code, the critical temperature in this case will be $T_c=0$. The system is therefore within the weakly topologically ordered phase at all finite temperature. Similarly the lifetime in all cases will be $O(L^2)$.

\textit{Coupling to bosons:-} The case of a toric code coupled to a system of hopping bosons has recently been considered \cite{fahut}. This has been shown to generate strong interactions between anyons and holes (the absence of an anyon on a plaquette or vertex). The resulting Hamiltonian has been shown to be well approximated by $H_0$, but with a value of $J$ that diverges with $L$.

There are two ways in which the code can be coupled to the bosons. For one of these the effective coupling scales as $J = O(L)$. The density of anyons is then exponentially suppressed by $L$, and so the thermal state is topologically ordered. In \cite{fahut} it is shown that the lifetime of the quantum memory scales as $O(\exp L)$ in this case. Note that similar behaviour is found for the model with anyon-hole repulsion in \cite{fabio,hutter,becker}, except that the interactions there lead to the stronger coupling $J = O(L^2)$, and the increased lifetime scaling $O(\exp L^2)$.

For the alternative coupling of the code to the bosons we find that $J = c \, T \ln(L/2)$. Here $c$ is an $O(1)$ constant that depends on the coupling between the code and the bosons, and $T$ is the temperature. Using Eq. (\ref{boltzmann}) we see that $\rho = (L/2)^{-c}$, and so Eq. (\ref{eq}) yields $\lambda = (L/2)^{c/2} \sqrt{\ln (1-\delta)}$. This gives a strongly topologically ordered phase for $c>2$ and a weakly topologically ordered one for $c<2$. In \cite{fahut} it is shown that the quantum memory is self-correcting in both regimes, with a lifetime of $L^{2c-2}$ for the former and $L^{c/2}$ for the latter. The boundary between the topological phases at equilibrium therefore again corresponds to a change in the scaling behaviour of the quantum memory.

\textit{Conclusions:-} Here we have studied equilibrium properties of the topological correlations in all interacting anyon models currently proposed for quantum memories. Comparisons are also made between this equilibrium behaviour and known results for the lifetime of the memories, a property of the dynamics during thermalization. It is found that the interactions are able to support topologically ordered phases at finite temperature. Two such phases are shown to exist, one of which is simply a finite temperature extension of the ground state phase, whereas the other is novel and distinct. For all models considered, being within one of these phases is a necessary and sufficient condition for a self-correcting quantum memory. The phase boundary between them corresponds to a change in the way the lifetime of the memory scales with system size. The novel `weakly topologically ordered' phase therefore appears to be an important and recurring feature of the kinds of models that support self-correcting quantum memories. The often more analytically and computationally tractable equilibrium case may therefore be used to determine whether a system is self-correcting when treatment of the dynamics is impractical.

\textit{Acknowledgements:-} The author would like to thank Adrian Hutter, Sergey Bravyi and Daniel Loss for inspiring discussions, and the Swiss NF, NCCR Nano and NCCR QSIT for support.

\end{document}